# LAVA WORLDS: EXOPLANET SURFACES

Marc-Antoine Fortin[1,2,3,*], Esteban Gazel[1,2], Lisa Kaltenegger[1,3,4], Megan E. Holycross[1,2]

[1] Carl Sagan Institute, Cornell University, Ithaca, NY, USA
[2] Earth and Atmospheric Science Department, Cornell University, Ithaca, NY, USA
[3] Cornell Center for Astrophysics and Planetary Science, Cornell University, Ithaca, NY, USA
[4] Astronomy Department, Cornell University, Ithaca, NY, USA
[*]Now at Science and Technology Division, Corning Incorporated, Corning, NY, USA

**ABSTRACT**

The recent first measurements of the reflection of the surface of a lava world provides an unprecedented opportunity to investigate different stages of rocky planet evolution. The spectral features of the surfaces of rocky lava world exoplanets give insights into their evolution, mantle composition and inner workings. However, no database exists yet that contains spectral reflectivity and emission of a wide range of potential exoplanet surface materials. Here we first synthesized 16 potential exoplanet surfaces, spanning a wide range of chemical compositions based on potential mantle material guided by the metallicity of different host stars. Then we measured their infrared reflection spectrum (2.5 - 28 µm, 350 - 4000 cm$^{-1}$), from which we can obtain their emission spectra and establish the link between the composition and a strong spectral feature at 8 µm, the Christiansen feature (CF). Our analysis suggests a new multi-component composition relationship with the CF, as well as a correlation with the silica content of the exoplanet mantle. We also report the mineralogies of our materials as possibilities for that of lava worlds. This database is a tool to aid in the interpretation of future spectra of lava worlds that will be collected by the James Webb Space Telescope and future missions.

**INTRODUCTION**

The discoveries of exoplanets with surface temperatures hot enough to produce melts ("lava worlds", e.g., Léger et al., 2009, Batalha et al., 2011; Demory et al., 2016) and the recent first measurements of the reflection of the surface of such a lava world (Kreidberg et al., 2019) provides an unprecedented opportunity to investigate different stages of rocky planet evolution. Furthermore, knowledge of the composition of the surfaces of these lava worlds is critical in understanding the composition and evolution of their interiors from petrologically and geodynamically accurate perspectives, as well as determining the potential of other planets in the same system or at a later stage to sustain life (e.g. Miguel et al. 2011; Chao et al., 2021).

The atmospheric and surficial compositions of planets are key in the determination of their habitability potential (e.g. Kaltenegger et al. 2017), and those factors are in turn directly controlled by planet-scale magmatic and recycling processes. During the Hadean, because of the heat generated by accretionary impact, even Earth was covered by a magma ocean (e.g. Elkins-Tanton, 2012). Little evidence remains of this period in the geologic record, but the compositional segregation and volcanic and tectonic activity that resulted past this magma ocean stage were likely important factors in the emergence of life on Earth (e.g. Dasgupta and Grewal, 2019).

With the recent launch of the James Webb Space Telescope (JWST), we will soon be capable of obtaining information about the surface composition of lava world exoplanets - both young rocky worlds that are initially hot like Earth was, as well as close-in rocky worlds - through visible to infrared measurements. JWST will be capable of measuring their thermal emissivity owing to the planets' high surface temperatures. However, no database of reflection and emission spectra data for the range of possible compositions of exoplanet surface materials exists yet.

Emissivity is a physical property that relates to reflectivity using Kirchhoff's approximation (1-reflection; Kirchhoff, 1860). Glasses and melts lack most infrared spectral features normally observed in rocky, crystal-rich materials because of their amorphous character. The Christiansen feature (CF; Henry, 1948), where reflectivity approaches zero and conversely, emissivity tends to unity has previously been used for remote sensing studies as a way to approximate the composition or temperature of Earth materials (e.g., Cooper et al., 2002; Pisello et al., 2019). The CF occurs at around 8 μm in silicate glasses and melts.

Volcanism on the surface of a planet is the mechanism that generates new crust from partial melts of the mantle. This crust, the surface of the planet, is of a different composition than the mantle source as it partially melts through various processes (e.g., Bowen, 1928; Grove and Brown, 2018). While on Earth, volcanism is mostly the result of processes related to plate tectonics or deep recycling via mantle plumes, volcanic materials can also result from other processes such as impact melting, volatile loss (e.g., degassing of a magma ocean), or the melting (Byrne and Bezada, 2020). The partial melts generated by such processes rise to the surface and may erupt to generate new crust. There are several potential sources of heat that spur the melting of planetary bodies, like stellar irradiation for close-in planets (e.g. Miguel et al. 2011; Chao et al., 2021 and references therein). Such melting influences the formation of crustal material on exoplanets.

Like on Earth, petrological tools can help us relate the surface of exoplanets (measurable) to their interior (shielded), but most current models of mineral equilibria and melt formation are mainly calibrated for solar compositions and are not accurate for the wide range of possible exoplanet composition s (Putirka and Rarick, 2019) (Fig. 1). While some progress has been made through recent experimental work on two potential exoplanet mantle compositions to examine the pressure-temperature conditions of mineral equilibria and melt formation (Brugman et al., 2021), no spectral data of observable features associated with such lava worlds exists.

Here, we report the first laboratory reflection measurements of potential lava world surfaces as well as a new compositional index, specifically tailored to exoplanet surface compositions, with the goal of informing upcoming observations by JWST, the ELTs and future missions.

## 1. METHODS
### 1.1 Petrological modeling and starting compositions

Recently, Putirka and Rarick (2019), modeled the bulk silicate planet (BSP) compositions of over 4000 hypothetical rocky planets from the composition of their stars, as reported in the Hypatia Catalog (Hinkel et al., 2014). They report primitive silicate mantles (bulk silicate planet, BSP, before the segregation of the crust) mostly comprised of olivine and/or orthopyroxenes, with possible variations depending on the partitioning of Fe during core formation. While many of their modeled BSP have Earth-like, peridotitic mantle compositions, pyroxenite planets are also important, with some compositions even reaching a silica-rich orthopyroxenite composition. These approximations of the composition of exoplanetary mantles provide a good starting point for the modeling of exoplanetary surface compositions.

We selected eight BSP compositions (as modeled by Putirka and Rarick 2019; 14684-30, 81262-30, 6856-50, 6856-30, 6856-10, 65356-30, 59639-30, 22907-30) as well as an average MORB composition (Gale et al., 2013), a chondritic composition (Chond; McDonough and Sun, 1995), and a peridotite (MM3; Baker and Stolper, 1994), spanning a wide range of potential exoplanet surface composition (see Fig. 1 and Supplementary Material Table 1). To more accurately represent the surface composition of lava world exoplanets, we modeled the composition of partial melts derived from these starting compositions as a two-stage process, the segregation of a crust derived from the BSP mantle and melting.

We first used rhyolite-MELTS (Gualda et al., 2012) to simulate crustal segregation during partial melting at a 1 atm pressure and an oxygen fugacity ($fO_2$) of about what is set by the quartz-fayalite-magnetite mineral oxygen buffer (QFM), as it is unlikely that the crustal composition of these objects is exactly their BSP, and much more likely that their surface is the result of some degree of partial melting of their mantle. We assumed melt fractions of 30% (F=0.3) for most compositions, but also modeled multiple melt fractions (F=1, 0.5, 0.3, 0.1) for MORB and 6856 to track potential compositional evolution of the surface from a given BSP. Additionally, we modeled F=0.1 for the peridotite (MM3) and 112774, for a total of 16 compositions (see modeled compositions plotted in Fig. 1). To synthesize these modeled compositions representing total and partial melting, we mixed the appropriate amounts of oxides ($SiO_2$, $TiO_2$, $Al_2O_3$, MgO, MnO) and carbonates ($CaCO_3$, $Na_2CO_3$, $K_2CO_3$) in an agate mortar and pestle under ethanol three times for ten minutes each. We decarbonated the mixtures at 1000 °C overnight, then crushed and remixed the ensuing material in an agate mortar and added FeO.

To simulate the conditions of an "airless planet" (e.g., near complete loss of atmosphere given the proximity to the star and/or low gravity), we equilibrated our compositions in controlled atmosphere Deltech vertical tube furnace in the experimental geochemistry lab at Cornell University. We placed each of the mixed and decarbonated compositions in a graphite crucible at 1310 ± 5 °C, soaking in a 100% CO atmosphere with an estimated $fO_2$ of about QFM-4 for one hour. The charges were then quenched in air and crushed for analysis with an integrating sphere, making sure to save a large chip of each composition, which we mounted in Epoxy and polished using SiC polishing paper and 1 µm $Al_2O_3$ powder for LA-ICP-MS and FTIR measurements (refer to section 2.2).

## 2.2 Analytical procedure

We analyzed each glass at the Cornell University Mass Spectrometry facility (CMaS) with an ESI NWR193UC excimer laser and an Agilent 8900 ICP-QQQ (ICP-MS/MS), operating in single-MS mode, for their major element composition. We processed the data following the method of Humayun et al. (2010), which provides major element processions and accuracies similar to that of electron microprobe analysis through standardization and calibration. We undertook the laser-ablation ICP-MS/MS measurements with 100 um spot sizes (except 50 µm for Chond-30 because of the small size of glass pools), 50% laser energy, 5 J/cm fluence, and Ar transport gas.

We collected IR spectra for each sample at room temperature using a Bruker Hyperion 2000 FTIR microscope in reflection mode, connected to a Bruker Vertex 80 at the Lava World Lab located at the Department of Earth and Atmospheric Sciences at Cornell University. We collected each spectrum using a liquid nitrogen-cooled MCT detector, KBr beam splitter, and a 15x IR objective using 128 scans from 4000 to 350 cm$^{-1}$ (4 to 28 µm). We obtained background spectra on a gold-coated reflector using the same parameters. We also used the same detector and parameters with a Bruker A 562-G integrating sphere on crushed powders of the materials, including the two compositions that did not produce a glassy material (112774-10 and 100963-30).

We collected Raman spectra of mineral phases of samples that did not fully quench to a glass using a WITec Alpha-300R Raman confocal microscope with a 532 nm laser set to 12 mW and a 100x objective. We also compared the spectra to the online RRUFF database (Lafuente et al., 2015) using the WITec software TrueMatch.

## 2. RESULTS

Compositions of synthesized glasses analyzed with LA-ICP-MS at Cornell University are reported in Supplementary Material Table 1. We note that because all compositions yielded crystals, the composition of the analyzed glasses differs slightly from the modeled compositions. Nevertheless, those synthesized compositions represent a wide range of possible exoplanet surface compositions. In total alkali-silica space (TAS), a commonly-used bulk rock classification system, the compositions span the entire range of Earth rocks: nephelinite, basalt, trachy basalt, phonotephrite, basaltic trachy andesite, basaltic andesite, phonolite, trachy andesite, andesite, trachyte, rhyolite. This classification is somewhat misleading as it is Earth-centric, and only considers a handful of elements. However, the selected BSP mantles are generally more Fe-rich than most Earth rocks, and represent compositions truly alien to Earth.

Fourteen compositions yielded glass, with two of those being only partial melts. Chond-30 and 14684-30 both yielded about 50% glass and 50% minerals and can thus be considered partial melts. The other two compositions, 112774-10 and 100963-30, did not produce glasses and were therefore not analyzed with the Hyperion microscope (see section 2.2), but were analyzed similarly with an integrating sphere. For these synthetic exoplanet surfaces, the mineral phases present are olivine, clinopyroxene, and magnetite identified by their mineral structure and confirmed by Raman spectroscopy. We report emission spectra, calculated from the measured reflection spectra from one-dimensional Hyperion measurements in Fig. 2. Note that the spectra obtained from the Hyperion and from the sphere are not identical and the CF is shifted between the data collected using the sphere and the one using the microscope. We attribute this shift to slightly different mineralogical modes (cf. Scudder et al., 2021). With the microscope, we specifically targeted areas with large pockets of glass, while for the sphere, we collected the reflected spectrum of the bulk composition, undoubtedly including a larger proportion of crystalline phases.

All spectra show a prominent maximum emissivity (minimum reflectivity) at about 8 μm where it is close to unity. This is the CF, one of the only prominent features in this suite of glassy materials. The CF is a result of the refractive index of a material being equal to that of the medium (i.e., backscattering minimum), and produces a sharp peak at that given wavelength (Logan et al., 1973). The other regions of the spectra are closely matched between different compositions, both with the set of spectra collected with the microscope (one-dimensional reflection) and the integrating sphere (diffuse reflectance). The CF is clearly shifted between the different samples, as expected. As previously reported in the literature, we find that the CF shifts to shorter wavelengths with increasing silica content of the glass (Fig. 3).

### 2.1 Spectral features

In vibrational spectroscopy, features are directly related to different bonds and their lengths (e.g., King and Larsen, 2013). While this is more straightforward for the case of minerals, with well-defined structures, glasses and melts generally exhibit much fewer and broader features. The correlation with silica in our samples is important, as most of the surface area of the samples consist of quenched silicate melts where Si-O bonds dominate, but the suite of samples varies in much more than just silica content and so does the diversity of the bonds in their amorphous framework.

The spectra of our more crystal-rich compositions (Chond-30 and 14684-30) show additional features at about 10 to 11 μm. These features appear to be explained by the presence of crystals in the bulk compositions that were also present in the analytical area of the FTIR, thus including spectral features from these phases. This is evident when comparing to the spectra of olivine (forsterite, RRUFF ID: R040018.1) and clinopyroxene (diopside, RRUFF ID: R040009.1) from the RRUFF online database (Lafuente et al., 2015), as shown in Fig. 3c.

## 2.2 Implications for the composition of the surfaces of lava worlds

The CF can be used to obtain a broad idea of surface composition of lava world exoplanets as it occurs away from the *reststrahlen* bands, a spectral range linked to the vibrations of bonds of tetrahedrally-coordinated oxides (e.g., $SiO_2$), where prominent crystalline phases may require complex deconvolution to decipher.

The SCFM index (Eq. 1; Walter and Salisbury, 1989) has been used to relate the structural properties of Earth natural glasses to their spectral features, such as the CF:

$$SCFM = \frac{SiO_2}{SiO_2+CaO+FeO+MgO} \ (wt.\%) \quad (1)$$

where each oxide is their mass proportion (e.g., wt.%). Figure 3a,b show the fit between our data and $SiO_2$ (wt.%) and the SCFM index. The SCFM was developed for Earth surface materials and proves a lackluster fit for our exoplanet compositions. Instead, we propose a new, empirical fit for exoplanetary surfaces:

$$CF\ (\mu m) = 11.1 \pm 0.4 - 3.9 \pm 0.4\ (X_{SiO_2}) - 5 \pm 1\ (X_{Al_2O_3}) - 2.4 \pm 0.6\ (X_{MgO}) - 2.4 \pm 0.5\ (X_{FeO}) \quad (2)$$

where $X$ is the mole fraction of the specified oxide. While both the SCFM and our new model are empirical fits, the new parametrization fits our data much better, and the use of mole fractions of oxides provides a better mechanism for the chemistry and structure of the quenched melts. We observe however, that an important point of weakness in both indexes is the lumping of ferric and ferrous irons together and that the effect of the speciation of iron in the glass or melt on the CF needs more work.

Additionally, if only the position of the CF is available, then the estimation of the silica content of the material remains the best estimation possible, as shown in Fig. 3a:

$$CF = 9.02 - 0.02\ SiO_2 \quad (3)$$

where $SiO_2$ is in wt.% and the CF in μm.

## 3. DISCUSSION

Our synthetic glasses not only provide information about the spectral signatures of lava worlds, but also additional information about their possible mineralogies. After reaching temperatures of 1310 ± 5 °C, and even after being quenched rapidly in air, compositions Chond-30 and 14684-30 showed a significant amount of crystals (60% and over). This implies that, for lava planets of such compositions to be fully molten, the surface temperature must be hotter than 1310 °C. Furthermore, this implies that the presence of both olivine and clinopyroxene may be common in future measurements of real lava worlds. Some evidence of these crystalline phases remains present near the CF at about 10 to 11 μm, as shown in Fig. 3 and discussed above. Hotter temperatures still are necessary to produce a melt from a silica-rich exoplanet with a BSP mantle dominated by orthopyroxene and olivine, or highly-silicic compositions (e.g., 100963-30, 112774-10). We therefore do not expect completely glassy material at the surface of planets of such compositions.

Lava planets could be completely molten, partially molten, or cooled and crystalline. As such, mixtures of glass and crystals could be readily identified on the surfaces of exoplanets from their emission spectrum. Coupled with its surface temperature, this would more readily allow recovering the mantle composition of a rocky body than for completely molten planets, as phase equilibria will further constrain the petrology of the surfaces and their relation to the BSP composition and the host star metallicity.

## 4. CONCLUSION

The first observation of the surface of a lava world has shown the critical lack of a database of reflection and emission spectra data for the range of possible compositions of exoplanet surface materials. Here we present the first spectra of 16 synthesized potential exoplanet surfaces, spanning a wide range of chemical compositions from 2.5 - 28 μm (350 - 4000 cm$^{-1}$). We also establish the link between the composition and a strong spectral feature at 8 μm, the Christiansen feature (CF).

The CF is the most prominent feature observed on glassy or partially glassy compositions of synthetic exoplanet surfaces between 2.5 and 28 μm. It provides important information not only about the silica content of lava world surfaces, but also their $Al_2O_3$, FeO, and MgO contents. As such, the CF is a promising spectral feature in the investigation of the surfaces of lava worlds and a valuable tool in further modeling and understanding their interior and evolution.

Because the surfaces of lava worlds could range from completely molten to completely crystalline, mechanical mixtures of glasses and crystals could potentially be identified from the spectral features of these bodies. This could help infer the composition of the mantle as phase equilibria would further constrain the possible petrologies and the relationship between their surface and BSP.

In the era of JWST, the exploration of lava planets will begin. Our database provides the first guide to a wide range of potential exoplanet surfaces--from surfaces that are completely molten, and partially molten, to cooled and crystalline ones, with mixtures of glass and crystals. It contains information on the mineralogies present in our experiments to inform the interpretation of future spectra. Our database provides a tool for observers to explore lava worlds.

## ACKNOWLEDGEMENTS

We would like to thank Dr. Emily First for valuable discussions on exoplanets and spectroscopy. This research was supported by the Heising-Simons Foundation (2019-1498).

# FIGURES

**Figure 1**: Selected compositions from Putirka and Rarick (2019; filled symbols) and their modeled partial melts (empty symbols) plotted on total alkali silica (TAS; a) and $SiO_2$-$FeO$+$MgO$-$Al_2O_3$+$CaO$+$Na_2O$ (b) diagrams.

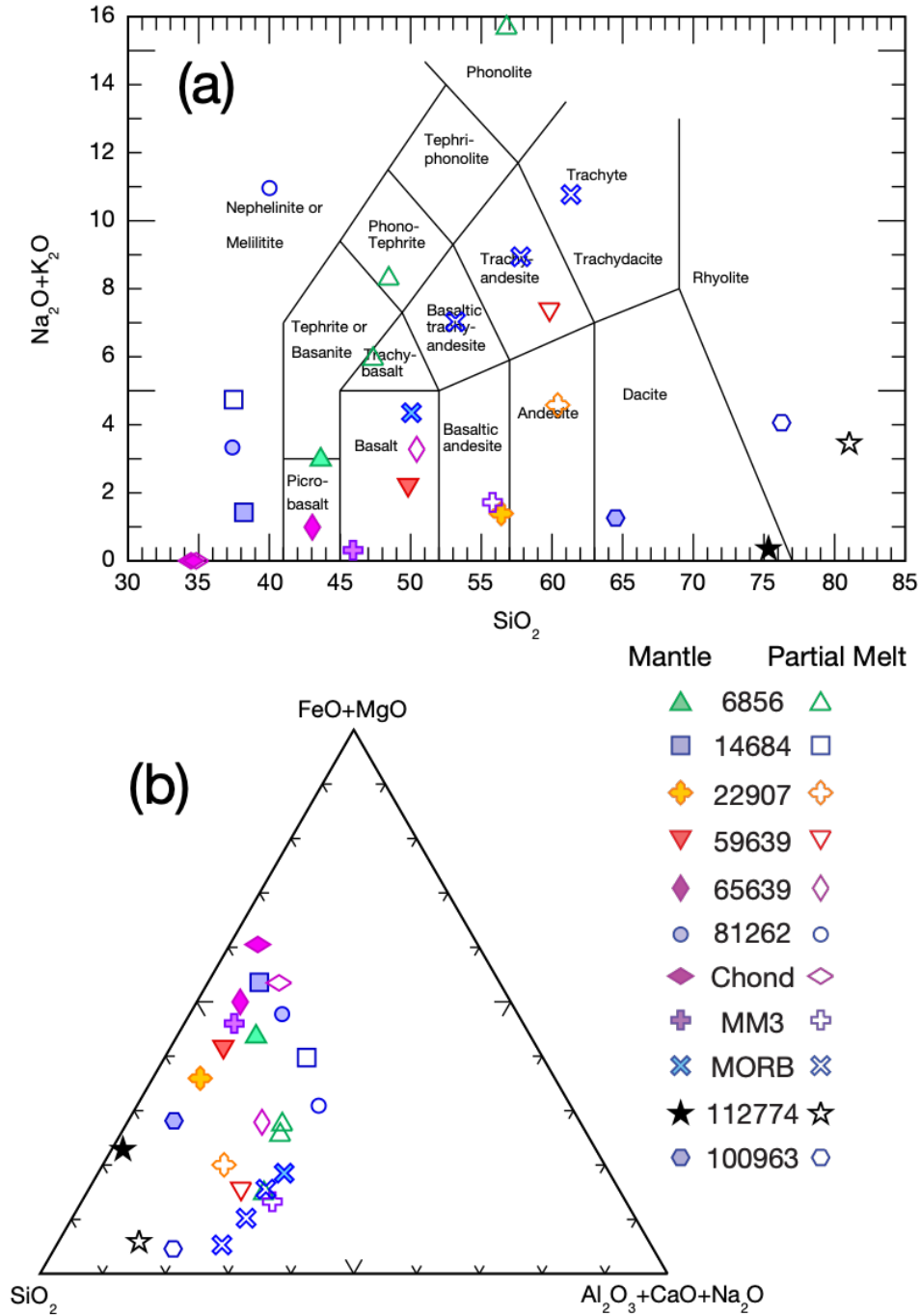

**Figure 2:** Calculated emission infrared spectra from measured reflection (1-R) of synthesized potential exoplanet surficial compositions. Acquired with a microscope FTIR (a) and a gold-coated integrating sphere (b). The crystalline features of the IR spectra of our more crystal-rich compositions (14684-30 and Chond-30) are compared to pure diopside and forsterite spectra in (c).

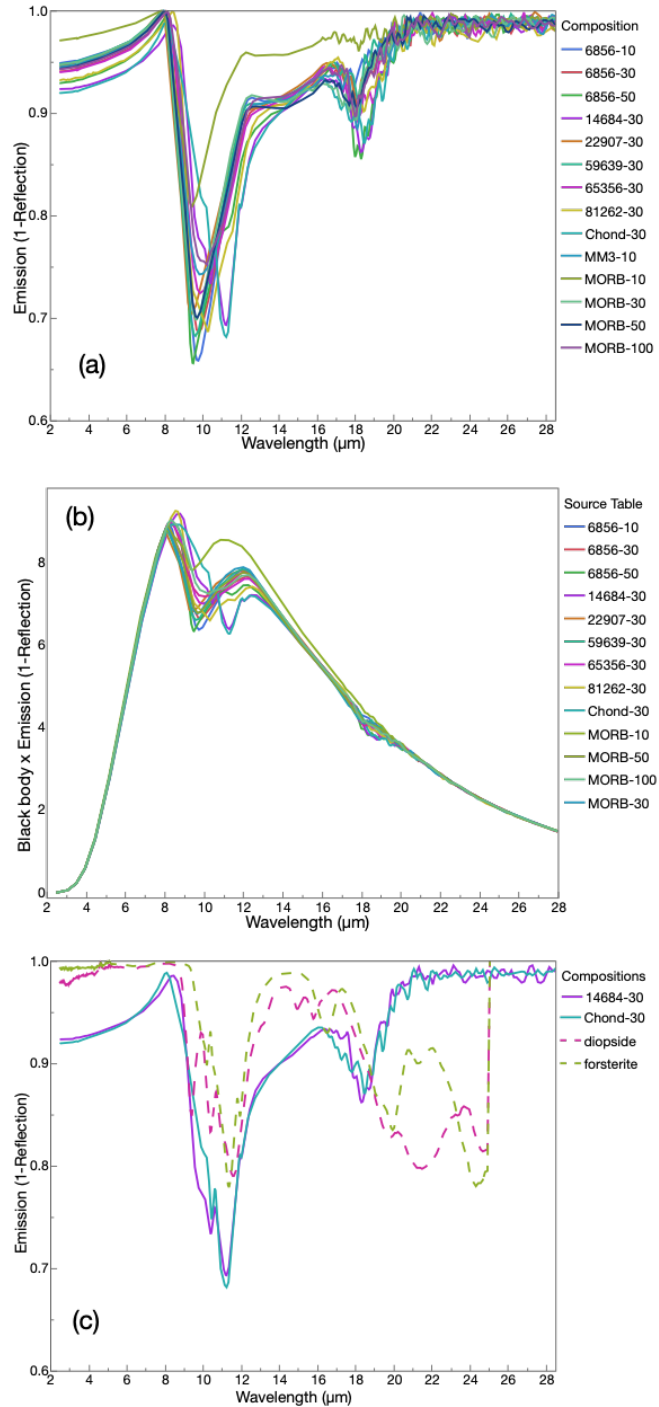

**Figure 3:** The location of the Christianssen Feature (CF) as a function of (a) $SiO_2$, (b) the SCFM index (Walter and Salisbury, 1989; Eq. 1), and (c) modeled using our new parameter (Eq. 2). Red data are obtained from the integrating sphere and blue data are from the Hyperion microscope. Dotted lines show linear and modeled fits to the data.

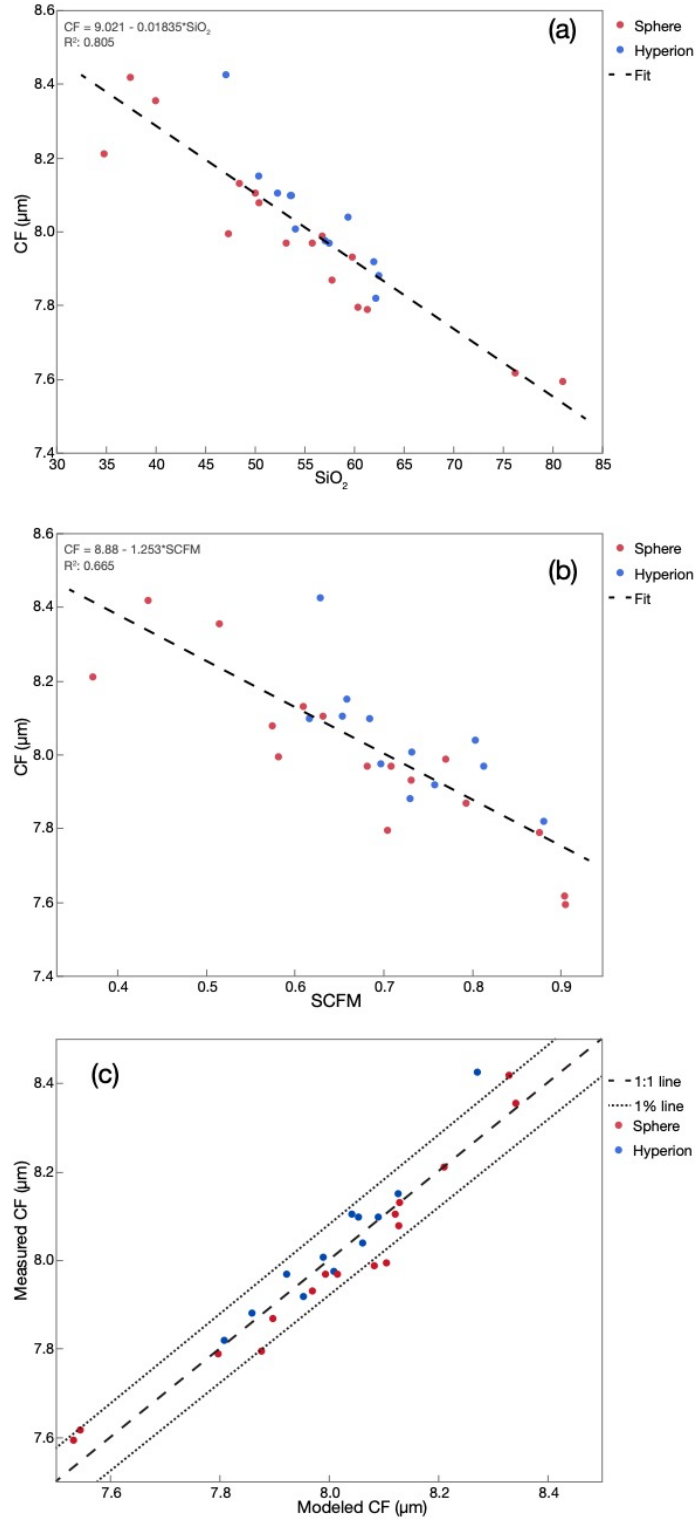

**SUPPLEMENTARY MATERIAL**

**Table 1:** Partial melt compositions modeled using Rhyolite-MELTS (Gualda et al., 2012) from selected compositions computed by Putirka and Rarick (2019).

| Composition | SiO2 | TiO2 | Al2O3 | FeO | MnO | MgO | CaO | Na2O | K2O |
|---|---|---|---|---|---|---|---|---|---|
| Chond-30 | 34.82 | 0.21 | 5.46 | 43.52 | 0.78 | 9.42 | 5.79 | 0.00001 | 0 |
| 14684-30 | 37.47 | 0.42 | 8.61 | 29.84 | 0 | 9.74 | 9.18 | 4.74 | 0 |
| 81262-30 | 40.01 | 0.32 | 11.02 | 24.37 | 0 | 6.43 | 6.89 | 10.96 | 0 |
| 6856-50 | 47.35 | 0.69 | 11.98 | 19.47 | 0 | 7.77 | 6.82 | 5.93 | 0 |
| 6856-30 | 48.45 | 1.04 | 11.2 | 20.29 | 0 | 4.95 | 5.78 | 8.29 | 0 |
| MORB-100 | 50.06 | 1.67 | 14.58 | 10.34 | 0.18 | 7.52 | 11.30 | 2.77 | 1.59 |
| 65356-30 | 50.43 | 0.5 | 8.47 | 18.96 | 0 | 8.76 | 9.6 | 3.28 | 0 |
| MORB-50 | 53.17 | 2.12 | 15.54 | 10.54 | 0.29 | 4.17 | 7.14 | 4.01 | 3.02 |
| MM3-10 | 55.8 | 0.97 | 15.48 | 3.73 | 0 | 9.41 | 12.9 | 1.72 | 0 |
| 6856-10 | 56.8 | 0.18 | 10.43 | 12.89 | 0 | 1.87 | 2.17 | 15.67 | 0 |
| MORB-30 | 57.79 | 1.88 | 15.97 | 6.69 | 0.37 | 2.84 | 5.52 | 4.36 | 4.58 |
| 59639-30 | 59.83 | 0.44 | 10.36 | 10.35 | 0 | 5.28 | 6.34 | 7.4 | 0 |
| 22907-30 | 60.4 | 0.39 | 9.31 | 14.82 | 0 | 5.13 | 5.36 | 4.58 | 0 |
| MORB-10 | 61.35 | 2.07 | 16.7 | 3.13 | 0.45 | 1.64 | 3.9 | 3.05 | 7.72 |
| 100963-30 | 76.25 | 0.29 | 11.35 | 2.79 | 0 | 1.79 | 3.47 | 4.06 | 0 |
| 112774-10 | 81.04 | 0.25 | 6.79 | 1.93 | 0 | 3.96 | 2.57 | 3.46 | 0 |

**Table 2:** LA-ICP-MS results from synthesized compositions in wt.%

| Sample | SiO2 | TiO2 | Al2O3 | MnO | MgO | FeO | CaO | Na2O | K2O |
|---|---|---|---|---|---|---|---|---|---|
| Chond-30 | 44.8 (8) | 0.26 (4) | 6.7 (7) | 0.73 (4) | 7 (2) | 33.4 (6) | 7 (1) | 0.08 (1) | 0.0032 (5) |
| 14684-30 | 44.1 (3) | 0.42 (2) | 9.8 (3) | 0.0286 (5) | 5.8 (2) | 24.1 (7) | 10.4 (6) | 5.2 (2) | 0.009 (2) |
| 81262-30 | 47.1 (3) | 0.352 (4) | 12.3 (1) | 0.0207 (4) | 6.49 (2) | 14 (5) | 7.34 (8) | 12.4 (1) | 0.0038 (1) |
| 6856-50 | 52.3 (6) | 0.71 (2) | 12.9 (2) | 0.019 (1) | 7.2 (1) | 13.3 (8) | 7.23 (2) | 6.28 (9) | 0.0036 (2) |
| 6856-30 | 54 (1) | 1.04 (2) | 12 (4) | 0.019 (1) | 4.9 (1) | 14.1 (6) | 5.7 (2) | 8.7 (2) | 0.0034 (3) |
| MORB-100 | 50.4 (5) | 1.85 (5) | 16.4 (6) | 0.027 (3) | 6.9 (3) | 7.2 (7) | 12 (3) | 3.3 (2) | 1.7 (1) |
| 65356-30 | 53.7 (2) | 0.525 (3) | 8.95 (2) | 0.0184 (2) | 8.93 (4) | 14.6 (1) | 9.87 (8) | 3.43 (2) | 0.0054 (3) |
| MORB-50 | 54 (1) | 2.24 (9) | 16.6 (5) | 0.316 (9) | 4.3 (2) | 7.7 (3) | 7.8 (4) | 4.17 (6) | 2.8 (2) |
| MM3-10 | 57.1 (5) | 0.95 (2) | 15.2 (1) | 0.0033 (2) | 9.27 (9) | 2.56 (7) | 12.9 (3) | 1.98 (2) | 0.0028 (1) |
| 6856-10 | 59.4 (7) | 0.181 (1) | 10.2 (2) | 0.0124 (9) | 1.96 (2) | 10.1 (7) | 2.4 (1) | 15.83 (2) | 0.0038 (5) |
| MORB-30 | 58 (1) | 2.2 (1) | 17 (1) | 0.4 (2) | 3 (1) | 3.8 (6) | 6.4 (3) | 4.7 (1) | 4.59 (8) |
| 59639-30 | 62 (7) | 0.451 (4) | 10.2 (2) | 0.0104 (5) | 5.3 (1) | 7.8 (6) | 6.7 (3) | 7.4 (2) | 0.0045 (6) |
| 22907-30 | 63 (1) | 0.41 (1) | 9.5 (2) | 0.0161 (7) | 5.46 (6) | 11.7 (9) | 5.9 (2) | 4.44 (2) | 0.008 (1) |
| MORB-10 | 62 (3) | 2.11 (7) | 16 (1) | 0.46 (5) | 1.6 (2) | 2.5 (5) | 4.3 (3) | 3.1 (3) | 7.6 (5) |

*Each data is the average of 3 individual measurements on the same compositions.
**Numbers in parentheses represent the last significant figures of standard deviation of the averages.